\documentclass[12pt]{iopart}
\usepackage[utf8]{inputenc}
\usepackage{csquotes}
\usepackage{graphicx}
\usepackage{float}

\usepackage{iopams}

\usepackage{perpage}
\usepackage{csquotes}
\usepackage{enumitem}
\usepackage{blindtext}
\usepackage{tensor}
\usepackage{soul}
\usepackage{url}
\usepackage{afterpage}
\usepackage{cite}
\usepackage{csvsimple}
\usepackage{pdflscape}
\usepackage{multicol}
\usepackage{multirow}
\usepackage{hyperref}
\usepackage{xcolor}
\usepackage{orcidlink}
\usepackage{caption}

\begin{document}

\title[Search and inference of GWs from MBHBs with PyCBC]{Adapting the PyCBC pipeline to find and infer the properties of gravitational waves from massive black hole binaries in LISA}

\author{Connor R. Weaving$^{1*}$ \orcidlink{0009-0008-2697-2998},
        Laura K. Nuttall$^1$ \orcidlink{0000-0002-8599-8791},
        Ian W. Harry$^1$ \orcidlink{0000-0002-5304-9372},
        Shichao Wu$^2$ \orcidlink{0000-0002-9188-5435} and
        Alexander Nitz$^{3,2}$ \orcidlink{0000-0002-1850-4587}
        }

\address{$^1$ University of Portsmouth, Portsmouth, PO1 3FX, United Kingdom}
\address{$^2$ Max-Planck-Institut f\"{u}r Gravitationsphysik (Albert-Einstein-Institut), D-30167 Hannover, Germany}
\address{$^3$ Department of Physics, Syracuse University, Syracuse NY 13244, USA}
\address{$^*$ Corresponding author}

\begin{abstract}
The Laser Interferometer Space Antenna (LISA), due for launch in the mid 2030s, is expected to observe gravitational waves (GW)s from merging massive black hole binaries (MBHB)s. These signals can last from days to months, depending on the masses of the black holes, and are expected to be observed with high signal to noise ratios (SNR)s out to high redshifts. We have adapted the PyCBC software package to enable a template bank search and inference of GWs from MBHBs. The pipeline is tested on the LISA data challenge (LDC)'s Challenge 2a (\enquote{Sangria}), which contains MBHBs and thousands of galactic binaries (GBs) in simulated instrumental LISA noise. Our search identifies all 6 MBHB signals with more than $92\%$ of the optimal SNR. The subsequent parameter inference step recovers the masses and spins within their $90\%$ confidence interval. Sky position parameters have 8 high likelihood modes which are recovered but often our posteriors favour the incorrect sky mode. We observe that the addition of GBs biases the parameter recovery of masses and spins away from the injected values, reinforcing the need for a global fit pipeline which will simultaneously fit the parameters of the GB signals before estimating the parameters of MBHBs.
\end{abstract}

\maketitle

\setcounter{page}{1}
\section{Introduction}

In preparation for the Laser Interferometer Space Antenna (LISA) \cite{1702_00786} launch in the mid 2030's, considerable efforts are being made to design and develop the data analysis tools needed to extract and examine the numerous different types of gravitational wave (GW) signals expected to be found in the LISA band. LISA will probe the mHz GW spectrum, allowing us to explore sources such as massive black hole binaries (MBHB) with total mass between $10^5-10^8M_\odot$ \cite{wise2023formation}; intermediate-mass black hole binaries (IMBHB) with total masses in the range $10^2-10^5M_\odot$ \cite{2303.00015}; extreme mass-ratio and intermediate mass-ratio inspirals (EMRI and IMRI) where coalescences have mass ratios of $10^{-6}-10^{-3}$ and $10^{-3}-10^{-1}$ respectively, with total masses in the range $10^{3}-10^{7}M_\odot$ \cite{Barack_2018, Gair_2017}, and more. Each of these systems, if detected and analysed properly, will help constrain and test a wide range of theories. For example, the effect of dark matter on a GW propagating from a MBHB or constraining the deviation from the Kerr metric using EMRIs (for example see Ref \cite{Barausse_2020}). Depending on the model of MBHB populations, the merger rate is estimated to be from O(2) - O(100) $yr^{-1}$ \cite{Li:2022fno, Li:2022rgm, Steinle:2023vxs, Fang:2022cso}.

In order to ensure that we can detect and analyse these signals, the LISA data challenge (LDC) \cite{lisa_data_challenge, baghi2022lisa} generates mock LISA data with various GW signals injected into simulated LISA noise. The series of challenges will eventually replicate all of the complexities that the true LISA data will contain, such as gaps in the data; noise artefacts found within the data that will affect the ability to infer a system's parameters; realistic LISA spacecraft orbits; a time varying estimate of the noise to account for longer, GW signals. All of these issues will need to be addressed before LISA data can be analysed. Several groups are working on different aspects of LISA data analysis and their work is highlighted in Refs \cite{Cornish:2005qw, Vallisneri:2008ye, Karnesis:2023ras, Bayle:2023qfo, Cornish:2003tz, Rubbo:2003ap, Petiteau:2008zz, Littenberg:2023xpl, Bayle:2022okx, Digman:2022jmp}.

In this paper, we present our pipeline which was developed to analyse the second LDC, referred to as \enquote{Sangria} \cite{sangria, ldc_manual}. The challenge consists of a training and blind data set where both the signals and the Gaussian instrumental noise are generated in the same way. We used the training data set to test and develop our pipeline as the number of signals with their injected parameters were given with the data. We then ran our analysis on the blind data, which contains an unknown number of signals. Our results for the blind data challenge were submitted to the LDC in October 2022. The blind challenge has now finished and the injected parameter values have been released publicly and can be accessed in Ref \cite{ldc_sangria_results}. 

There are two types of GW signals injected into Sangria. The first type of signal is galactic white dwarf binaries (GBs) \cite{Nelemans_2001} which are expected to be the most abundant source of GW signals present in LISA data. The accumulation of these signals forms a \enquote{confusion noise} which dominates the $0.5-3$ mHz range of the frequency spectrum. A subset of these are known as verification binaries (VBs) \cite{finch2022identifying}, which are signals that are known in advance due to electromagnetic observations. The second type of signal in the data are MBHBs. These are generated using the IMRPhenomD waveform model \cite{imrphenomd_1, imrphenomd_2}. IMRPhenomD produces two GW polarizations which encode the effect a GW has on the relative distance between particles, the derivation of which is well known and can be found in Ref \cite{maggiore2008gravitational}. These polarizations are then projected onto the LISA constellation to produce an arm response. The arm response from LISA is known to be dominated by laser frequency noise and therefore Time Delay Interferometry (TDI) \cite{PhysRevD.68.061303, Tinto_2004} techniques must be used in order to extract GW information from the raw data streams. The Michelson TDI variables X, Y and Z are provided in this data challenge.

\begin{figure}[t!]
	\centering
	\includegraphics[width=0.9\linewidth]{"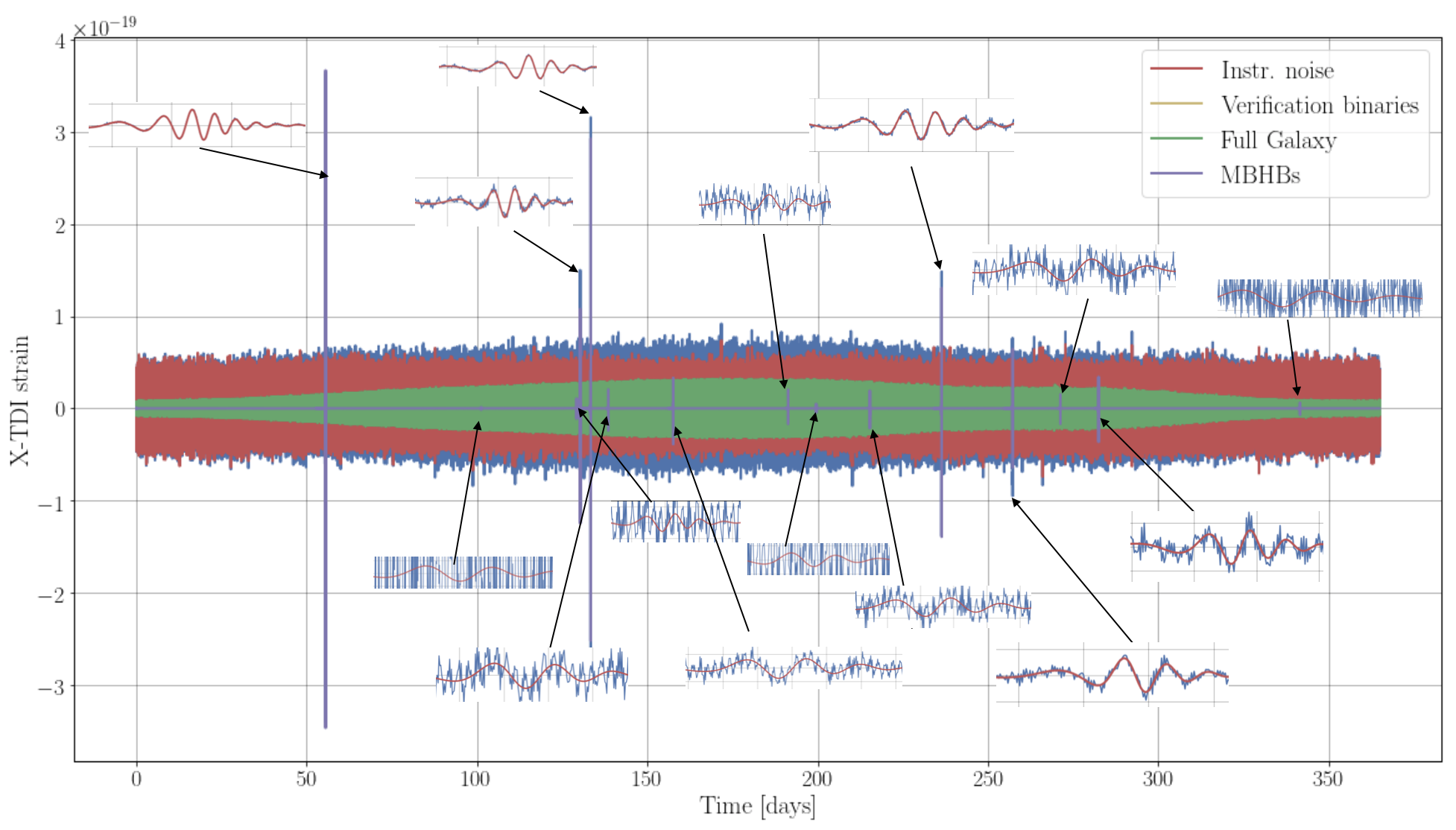"}
	\caption{Timeseries of X TDI variable from the Sangria training data set. The data consists of a year's worth of data sampled at 5s, with 15 MBHB signals injected into the data. This data also contains Gaussian instrumental noise and 30 million GBs, of which 17 are VBs. This image was taken from the LDC website for the Sangria challenge in Ref \cite{sangria}.}
	\label{fig:sangria_td}
\end{figure}

The data set spans a year's worth of data sampled at 5 seconds, which is the expected sample rate for LISA. Figure \ref{fig:sangria_td} shows a time series of the X TDI data stream of the Sangria training data set. An overview of previous LDCs can be found in Refs \cite{MLDC, MockLISADataChallengeTaskForce:2009wir, Arnaud:2007jy}.

The tools used in our analysis have been developed within PyCBC \cite{PyCBC}. PyCBC is an open source GW data analysis software package and is one of the main pipelines used to detect and characterise compact binary mergers from LIGO, Virgo and KAGRA (LVK) detectors \cite{Usman_2016, Biwer:2018osg, alex_nitz_2022_6324278,LIGOScientific:2016vbw}. PyCBC has been used to detect and analyse $\mathcal{O}(100)$ GW signals from compact binary mergers to date, an overview of which can be found in Refs \cite{PhysRevX.11.021053, theligoscientificcollaboration2021gwtc3, nitz20224ogc}. We have adapted the techniques developed for LVK searches and applied them to LISA-like signals, in this case MBHBs. 

It is important to note that there is far more work to do to achieve a reliable analysis of MBHBs in realistic conditions. The tools made available through this paper are a first attempt at the bigger picture solution, which will require many more iterations before being considered \enquote{production ready}.

The paper is laid out as follows. In Section 2, we introduce PyCBC and the components which make up our pipeline. We give the basic details needed to understand how PyCBC approaches detecting and recovering GW system parameters. In Section 3, we outline how we have adapted PyCBC in order to detect and infer MBHBs. In Section 4, we present the results our search and each stage of inference leading to the analysis of the blind Sangria data. Finally in Section 5 we discuss the challenges in developing these tools, future work and our conclusions.

\section{PyCBC: Search and inference}

We can split the PyCBC analysis pipeline up into two distinct sections, the search and the inference. The search portion of the pipeline is responsible for identifying candidate GW signals within the data, and the inference portion returns a statistical estimate of the parameters of the GW signals identified. Below we outline which tools within PyCBC have been used and adapted for MBHB signals. 

\subsection{The search}

The main component of the search is the application of matched filtering. The matched filter signal-to-noise ratio (SNR) is designed to indicate when a particular signal we are searching for could be present in data \cite{mf_ref}. For data in the form $d(t) = h(t) + n(t)$, where $h(t)$ is a modelled signal we are looking for and $n(t)$ is the detector noise, we can choose a linear filter, $F(t)$, such that the output SNR is maximised for the signal $h(t)$. We can represent this filtering acting in the time or frequency domain, as shown in Equation \ref{crosscorr and SNR},

\begin{equation}
    \label{crosscorr and SNR}
    \phi =  \int^{\infty}_{-\infty} F(t)d(t)dt = \int^{\infty}_{-\infty} \tilde{F}^{*}(f)\tilde{d}(f)df 
\end{equation}

\noindent
where tilde represents the Fourier transform. The optimal SNR defined as $\rho_{opt} = S/N$, where S is the value of $\phi$ where a signal $h(t)$ is present, and N is the root mean squared value of $\phi$ with no signal.

If we assume the noise to be Gaussian and stationary, we can define the the one-sided power spectral density (PSD) via $S_n$ in Equation \ref{psd},

\begin{equation}
    \label{psd}
    \langle\tilde{n}(f)\tilde{n}^{*}(f^{'})\rangle = \frac{1}{2}S_n(|f|)\delta(f - f^{'}). 
\end{equation}

\noindent
where $f\geq0$ and the square brackets represents the ensemble average. By introducing the inner product of two real time series $a(t)$ and $b(t)$ in Equation \ref{inner_prod1}

\begin{equation}
    \label{inner_prod1}
    (a|b) = 4 \Re \int^{\infty}_{0} \frac{\tilde{a}^{*}(f) \tilde{b}(f)}{S_{n}(f)}df,
\end{equation}

\noindent
we can express the optimal SNR as 

\begin{equation}
    \label{opt snr}
    \rho_{opt} = \frac{(u|h)}{(u|u)^{1/2}}, \quad \textrm{where} \quad \tilde{u}(f) = \frac{1}{2}\tilde{F}(f)S_n(f).
\end{equation}

\noindent
In order to maximise $\rho_{opt}$ it is clear that $\tilde{F}(f)$ should be proportional to $h(t)$

\begin{equation}
    \tilde{F}(f) = A \frac{\tilde{h}(f)}{S_n(f)},
\end{equation}

\noindent
where A is some arbitrary constant. This results in the optimal SNR taking the form

\begin{equation}
    \rho_{opt} = (h|h)^{1/2}.
\end{equation}

\noindent
In order to filter the data for a given signal, we would compute the matched filter SNR in Equation \ref{mf snr} which is distinct from $\rho_{opt}$

\begin{equation}
    \label{mf snr}
    \rho_{mf} = \frac{(d|h)}{(h|h)^{1/2}}.
\end{equation}

To generate an SNR timeseries, we can inverse Fourier transform the matched filter as shown in Equation \ref{mf ts snr},

\begin{equation}
    \label{mf ts snr}
    (d|h)(t) = \int^{\infty}_{-\infty} \frac{\tilde{d}(f)|\tilde{h}(f)|^{*}}{S_n(f)}\exp[-2\pi i f t]df,
\end{equation}

\noindent
where $t$ is time. Equation \ref{mf ts snr} shifts the waveform $h(t)$ to the defined time. By computing across the entire time of the data, the SNR timeseries returned indicates at which times the signal $h(t)$ is likely to be present by peaks in the SNR timeseries. 

We use a model for signals, $h(t)$, which we will now refer to as templates, that represents the response of a detector given a GW passing through the instrument. PyCBC can project the GW polarizations produced by varying waveform models onto different ground-based detectors to return an arm response. For some GW emitting systems, many parameters are required to define the full signal response. Given such a broad parameter space of values to explore, we require a large number of templates to search the data in order to potentially detect a GW. PyCBC generates template banks which can contain $\mathcal{O}(100,000)$ templates for a LVK-like search to explore the parameters which affect GW emission from compact binary coalescing systems. These parameters are split into intrinsic, which define the masses of both compact objects and their spins, and extrinsic, which describes the system's sky position, the luminosity distance to the binary and the orientation of the binary relative to the detector \cite{Allen:2005fk}. Details of the various methods used to generate template banks can be found in Refs \cite{Owen:1995tm, Owen:1998dk, Babak:2006ty, Keppel:2013yia, Privitera:2013xza}. 

When generating a template bank, the match, as defined in Equation \ref{match}, is computed between templates.

\begin{equation}
    \label{match}
    \mathcal{M}(h_{1}, h_{2}) = \mathrm{max}_{\phi_{c}, t_{c}} \left[ (h_1 | h_2) \right],
\end{equation}

\noindent
given two templates $h_1$ and $h_2$, the match tells us how similar to each other these templates are whilst maximising over the coalescence time $t_c$ and the phase at coalescence $\phi_c$. Templates are first normalised such that the inner product gives $(h_1|h_1) = 1$. Using the normalised templates in the match returns a number in the range [0,1] where 1 states that the two normalised templates are identical.

The matched filter is then applied to all templates and the data we are searching over, giving an SNR timeseries per template. Each timeseries is then passed to a clustering algorithm, to return the highest SNR template per time in the data. A trigger is defined by a peak in the SNR timeseries reaching, or surpassing, a certain threshold. Additional steps are then applied to search for coincidences over multiple detectors and calculate their significance \cite{Usman_2016, Davies:2020tsx}.

\subsection{The inference}

PyCBC can infer signal parameters through sampling the posterior probability defined by Bayes theorem

\begin{equation}
    \label{bayes}
    p(\underline{\theta} | d) = \frac{\mathcal{L}(d | \underline{\theta}) p(\underline{\theta})}{p(d)}
\end{equation}

\noindent
where $\mathcal{L}(d | \underline{\theta})$ is the likelihood of the data $d$ given the parameters $\underline{\theta}$, $p(\underline{\theta})$ is the prior of the parameters and $p(d)$ is the evidence and $\underline{\theta}$ represents an integer $N$ number of parameters $\underline{\theta} = \{\theta_1,...,\theta_N\}$. The main component of Equation \ref{bayes} to compute is the likelihood which under the assumptions of Gaussianity and stationarity takes the form:

\begin{equation}
    \log\mathcal{L}(d, \underline{\theta}) \propto (d|h_\theta) - \frac{1}{2}(h_\theta|h_\theta) -\frac{1}{2}(d|d),
\end{equation}

\noindent
where $h_\theta \equiv h(\underline{\theta})$.

We can speed up likelihood calculations by enforcing some assumptions about our GW. 

\begin{equation}
    \label{h_model}
    \tilde{h}(f; \underline{\theta}, \phi) = A(f, \underline{\theta})\exp[i(\Psi(f, \underline{\theta}) + \phi)].
\end{equation}

\noindent
We use IMRPhenomD (2,2) mode only signals, and express our GW strain, $\tilde{h}(f)$, as a product of smoothly varying amplitude and phase factors, where $f$ is frequency, $A(f,\underline{\theta})$ is the amplitude of the GW, $\Psi$ is the phase of the waveform and $\phi$ is some arbitrary phase constant.

With the GW strain expressed in this form, we use the application of heterodyning (sometimes referred to as relative binning) which assumes a smooth signal which varies in amplitude and phase. Heterodyning is a signal processing technique used to separate data into slowly and rapidly varying frequency components. Our inference application of heterodyning allows us to greatly speed up likelihood calculations by first finding a high likelihood template with our template bank search. The corresponding template is then used as a reference to explore the region of high likelihood. A derivation of this method applied to GWs can be found in Refs \cite{hetero_likelihood, relbin, Cornish:2021lje}.  

In order to explore the parameter space, there are several samplers in PyCBC which implement different methods, including Markov Chain Monte Carlo (MCMC) and Nested Sampling. An overview of PyCBC's inference setup can be found in Ref \cite{Biwer_2019}.

\section{Adapting PyCBC's search and inference for MBHBs}
\label{method}

\subsection{Templates}

In order to perform a matched filter search, we need to be able to generate templates that represent LISA's arm response given a GW signal passing through the instrument. The data we are searching over is given in terms of TDI variables; we therefore must put the templates in this format.

A vital part of this development is the use and implementation of the open source software BBHx waveform generator \cite{Katz_2020hku,Katz:2021uax,michael_katz_2021,Marsat:2018oam,Marsat_2021,London2018,Husa2016,Khan2016}, which we use to generate the expected strain from MBHB systems. BBHx generates the GW polarizations, projects them onto LISA to form an arm response and then creates the TDI channels. BBHx implements the IMRPhenomHM \cite{PhysRevLett.120.161102, abbott}, an aligned spin waveform model which produces GW polarizations based on IMRPhenomD, but includes other sub-dominant modes. In this work, we set the IMRPhenomHM to generate just the (2,2) mode in order to match the Sangria convention. We transform both the data and the TDI templates into noise independent data streams referred to as A, E and T \cite{tinto-tdi} via the transforms in Equation \ref{TDI variables}.

\begin{equation}
    \label{TDI variables}
    \fl A = (Z - X)/\sqrt{2},\quad
    E = (X -2Y + Z)/\sqrt{6},\quad
    T = (X + Y + Z)/\sqrt{3}.
\end{equation}

The data streams A and E have equal SNR across the entire LISA frequency band. This is not true for the T TDI stream as at lower frequencies the SNR contribution from T does not compete with A and E. However, T does contribute SNRs that can be greater than A or E at higher frequencies in the LISA band and is therefore included in the inference analysis \cite{Tinto_2005}. BBHx generates templates A, E and T in the same convention as in Equation \ref{TDI variables}.

\subsection{Template bank}

This work explores whether a template bank based search is a viable approach to finding MBHBs in LISA data, so to reduce complexity we only perform the search on the A TDI stream. We generated a template bank for the A TDI data stream, which was used for the filtering stage of the search on just the A TDI data. As the A and E TDI streams have equal SNR across the LISA frequency band, searching through just the A or just E data streams would be equivalent. We do lose power when searching over one instead of both of the TDI streams, but due to the SNRs of these signals being so large, we find that the information lost is negligible. Due to the SNRs of MBHBs being so large, we populated a less dense template bank when compared with LIGO template bank searches to balance accuracy of the signal trigger time versus computation cost of producing the bank.

For the bank generation we use stochastic placement \cite{Babak:2008rb, Harry:2009ea, Ajith:2012mn}. Specifically we use the PyCBC algorithm \enquote{brute bank} with the parameters defined in Table \ref{MBHB param ranges} and a minimal match of 0.9 which resulted in a bank size of 50 templates. The minimal match defines how densely the parameter space is sampled. For example, with a minimal match of 0.9, if any newly generated template matches with any other template in the bank above the minimal match, it is rejected and another template is generated. The process repeats until a pre defined number of templates have been rejected from the bank placement.

\begin{table}[t!]
    \begin{center}
        \begin{tabular}{||l|c|c||} 
             \hline
             Parameter & Lower Bound & Upper Bound \\ [0.5ex] 
             \hline\hline
             $M_T$ ($M_\odot$) & $2\times10^5$ & $2\times10^7$ \\ 
             \hline
             $q$ & 1 & 4 \\
             \hline
             $a_1$, $a_2$ & -0.99 & 0.99 \\
             \hline
             $\beta$ (rad) & $-\pi/2$ & $\pi / 2$ \\
             \hline
             $\lambda$ (rad) & 0 & $2\pi$ \\
             \hline
             $\psi$, $\iota$ (rad) & 0 & $\pi$ \\ [1ex]
             \hline
        \end{tabular}
    \end{center}
    \caption{\label{MBHB param ranges} The parameter ranges used to generate the template bank. This covers the full range of potential systems that are defined within the conventions of the Sangria challenge. $M_T$ is detector frame total mass of the black holes; $q$ is mass ratio where $m_1 > m_2$ and $m_1$ is the primary mass and $m_2$ the secondary; $a_1$ and $a_2$ are the aligned spins of the two black holes; $\beta$ is ecliptic latitude; $\lambda$ is ecliptic longitude; $\psi$ is polarization angle; $\iota$ is inclination.}
\end{table}

The luminosity distance $D_L$ is marginalised over in the matched filter calculations and is set to an arbitrary value when generating templates. The time of coalescence $t_c$ is sampled with one month from the beginning and end of the year removed. This is to ensure no edge corruption artefacts are present in the data when performing matched filtering, which could cause false triggers. 

When computing the match, maximising Equation \ref{match} over time of coalescence $t_c$ implies that we do not take into consideration the varying response function from LISA over time. LISA's response function varies over time due to the antennas sky position changing within the duration of the MBHB signal. However, because the majority of the SNR from merging MBHBs is within the final $\mathcal{O}(hour)$, and these systems have SNRs $\mathcal{O}(1000)$, we are safe to assume LISA is approximately stationary during this time period. 

As $t_c$ is directly related to LISA's orbital position in the case of BBHx, we ensure that the template bank can generate templates at different orbital positions by varying $t_c$. The actual recovery of $t_c$ is done via the filter output time series from the inner product, shown in Equation \ref{mf ts snr}, as $t_c$ is not an explicit template bank parameter.

\subsection{Sampling, priors and sky folding}

We sample the heterodyne likelihood using the Nested Sampling tool Dynesty \cite{Speagle_2020}. Table \ref{priors table} outlines the parameters used with their corresponding distributions in the inference. $M_c$ is the chirp mass of the system \cite{chirp}.

\begin{table}[t!]
    \begin{center}
        \begin{tabular}{||l|c|c||} 
             \hline
             Parameter & Lower Bound & Upper Bound \\ [0.5ex] 
             \hline\hline
             $M_c$ ($M_\odot$) & 87055 & 8705505 \\
             \hline
             $q$ & 1 & 4 \\
             \hline
             $a_1$, $a_2$ & -0.99 & 0.99 \\
             \hline
             $t_c$ (s)& $t_{ref}$ - 2 days & $t_{ref}$ + 2 days \\
             \hline
             $\beta$, $\lambda$, $\psi$ (rad) & 0 & $\pi / 2$ \\
             \hline
             $\iota$ (rad) & 0 & $\pi$ \\
             \hline
             $D_L$ (Mpc) & $10^3$ & $10^6$ \\ [1ex]
             \hline
        \end{tabular}
        \caption{A list of all parameters and their ranges given to the prior, $p(\underline{\theta})$. $M_c$, $q$, $a_1$, $a_2$, $t_c$, $\psi$ and $D_L$ are distributed uniformly. $\beta$, $\lambda$ and $\iota$ are distributed isotropically. $t_{ref}$ is the coalescence time found in the search.}
        \label{priors table}
    \end{center}
\end{table}

The importance of multi-modality and degeneracies when considering extrinsic parameters are summarised in detail in Ref \cite{Marsat_2021} and are vital to understand for our application of sampling we introduce here. Considering a (2,2) mode signal within the LISA frame, and applying the frozen low-f approximation which neglects both time and frequency dependence in the instrument response, there are 8 exact degeneracies for sky location. Four longitudinal at $\{\lambda_L + \frac{\pi}{2}(0,1,2,3), \psi_L + \frac{\pi}{2}(0,1,2,3)\}$ and two latitudinal at $\{\pm\beta_L, \pm\cos{\iota_L}, \pm\cos{\psi_L}\}$, where the subscript L represents the extrinsic parameters in the LISA reference frame.


\begin{equation}
    \label{r mode}
    \eqalign{\lambda^{(r)}_{L} &= \lambda_L, \\
    \beta^{(r)}_L &= - \beta_L, \\
    \psi^{(r)}_L &= \pi - \psi_L, \\
    \iota^{(r)}_L &= \pi - \iota_L, \\
    \phi^{(r)}_L &= \phi_c,}
\end{equation}


The BBHx waveform model that we use in this work, uses the frozen approximation and frequency-dependent instrument response which neglects LISA's motion. This breaks the 8-fold degeneracy into an approximate 2-fold degeneracy in the reflected mode shown in Equation \ref{r mode}. 

Instead of sampling the entire parameter space, we restrict our sky location parameters to one of eight octants of the sky. After drawing a sample from our defined octant, we compute the likelihood at the other 7 modes. We marginalise over the 8 sky modes by summation during the inference process. We then normalise them to sum to one. A post-processing step is then implemented to “unfold” the sky positions by randomly selecting one of the 8 sky modes based on the likelihood of each. This effectively samples the entire parameter space without having to sample other parts of the sky as we compute the likelihood directly at the analytical location of the other sky modes. 

All samples for inference are taken within the LISA reference frame, not the solar system barycentre (SSB) frame, even though the injected parameters in Sangria are all with reference to the SSB frame. The techniques described above for sampling the sky position are used in every inference run in this work.

\subsection{PSDs}

When GBs are included in the data, we assume them to be part of the noise $n(t)$. As there are several thousands of these signals of varying amplitude and phase, we can justify their accumulation to be Gaussian (but not stationary) by the central limit theorem. When constructing the PSDs for all three TDI variables, we implemented Welch's method over 31 segments of the year's worth of data. This gave frequency bins with a resolution of $10^{-6}$ Hz, which were narrow enough to model the structure of the PSD accurately. The same number of segments is used to generate all other PSDs for the different inference stages and the search. We generate a PSD per TDI channel at each inference stage.

By constructing the PSDs in this way we have ignored the fact that the PSDs are time-dependent and therefore cannot be stationary. This is also true in the case where the GBs are not included in the noise due to LISA's motion across the sky. We prioritised the LISA high-frequency information over the entire year's worth of data by reading all of the data and using 31 segments instead of a quasi-stationary PSD which is estimated on smaller chunks of data.

\begin{figure}[t!]
    \hspace{-1cm}
	\includegraphics[width=1.2\linewidth]{"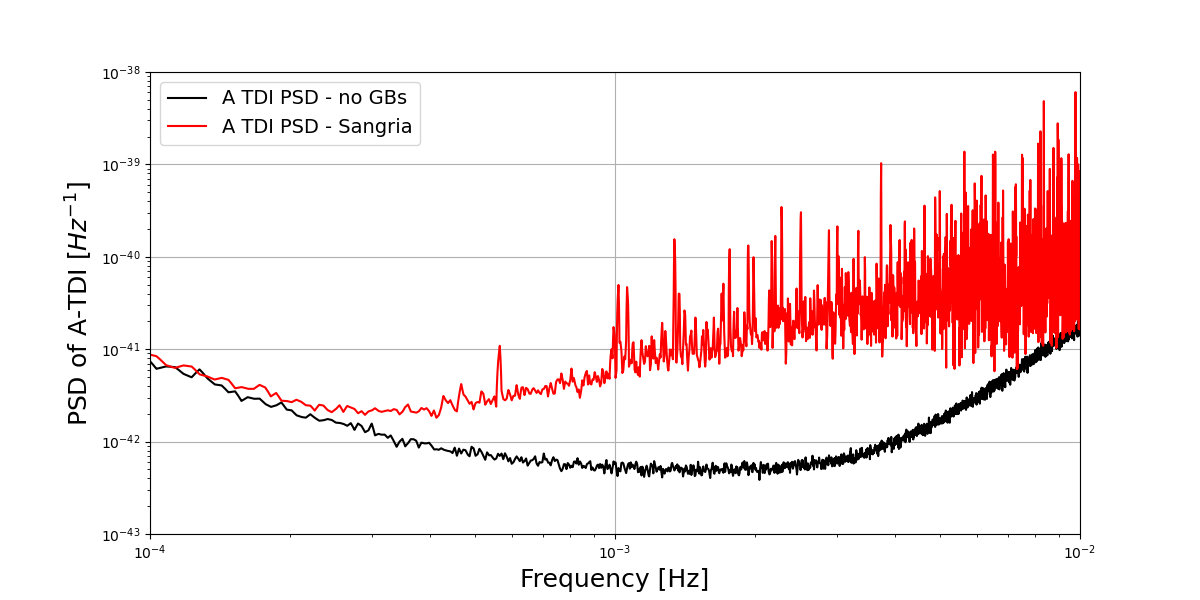"}.
    \vspace{-1cm}
	\caption{A figure showing the different PSDs we use in the results section. The red curve is the A TDI PSD of the Sangria data and the black curve is the A TDI PSD of Sangria without GBs. }
	\label{fig:psd}
\end{figure}

\section{Results}

In this section we present the results of our search and multiple inference stages. The associated data release which contains all of the code and instructions on how to reproduce the results presented can be found in Ref \cite{git_repo}. It contains the results for all analysis of the search and inference in all of their various stages. Documentation for ongoing developments in PyCBC for LISA related searches and inference can be found in Ref \cite{lisa_pe}.

\subsection{Search results}

To quantify how effective the template bank search was, we computed the match between the highest SNR template per trigger against the injected waveform parameters. All matches are greater than 0.92 and are presented in Table 3 for each candidate.

\begin{table}[t!]
    \label{Match_table}
    \centering
    \begin{tabular}{||c|c||}
        \hline
        Signal index & Match \\
        \hline\hline
        0 & 0.94510 \\
        \hline
        1 & 0.92666 \\
        \hline
        2 & 0.97344 \\
        \hline
        3 & 0.94091 \\
        \hline
        4 & 0.94562 \\
        \hline
        5 & 0.92439 \\
        \hline
    \end{tabular}
    \captionof{table}{The match defined in Equation \ref{match} between the highest SNR template per trigger and the waveforms injected value of the template bank search over the Sangria blind data.}
\end{table}

The template parameters from the search and the corresponding injected waveform parameters for signal 0 can be found in Table 4. We attempted to produce template banks at higher minimal matches but the time to complete was $\mathcal{O}(days)$ instead of $\mathcal{O}(minutes)$ in the 0.9 minimal match case. A template bank search as we have applied it worked exactly how we needed it when considering the balance of computational cost verses accuracy of the returned template.

\begin{table}[t!]
    \label{temp_bank_diff}
    \hspace{4cm}
    \begin{tabular}{||l|c|c||}
        \hline
        \bfseries Parameters & \bfseries Injected & \bfseries Template
        \begin{filecontents*}{"diff_table_search0_2.csv"}
Parameter,Injected,Recovered
m_1~(M_\odot),1.76\times10^{6},2.02\times10^{6}
m_2~(M_\odot),1.30\times10^{6},7.20\times10^{5}
a_1,0.51,0.27
a_2,0.14,-0.09
t_c~(s),6712124,6712065
\iota~(rad),1.31,2.27
\psi~(rad),1.06,5.51
\beta~(rad),-0.35,0.52
\lambda~(rad),2.31,2.78
\phi_{c}~(rad),4.23,3.14
\end{filecontents*}
\csvreader{"diff_table_search0_2.csv"}{}
        {\\\hline$\csvcoli$&$\csvcolii$&$\csvcoliii$}
        \\\hline
    \end{tabular}
    \caption{Comparison with template parameters found from the search verses the injected waveform parameters of signal 0 in the Sangria blind data.}
\end{table}

\subsection{Inference results}
\label{inf results}

In this section, we present our results from the inference. All of the runs have the reference parameters for the heterodyne likelihood set to the injected parameter values. Each inference run uses all three TDI data streams. Running the heterodyne likelihood with the reference values found in the search gave no significant bias when comparing to the reference parameters being set to the injected signal values.

When building the inference tools for this challenge, there were several stages of additional complexities which were successively built upon to ensure that each step performed as expected. We performed inference with increasing complexity as follows:

\begin{itemize}
    \item Injections with BBHx in zero noise.
    \item Injections with BBHx in Sangria training noise without GBs.
    \item Injections with BBHx in Sangria training noise with Sangria training GBs.
    \item Sangria blind data.
\end{itemize}

\begin{figure}[t!]
	\centering
	\includegraphics[width=0.7\linewidth]{"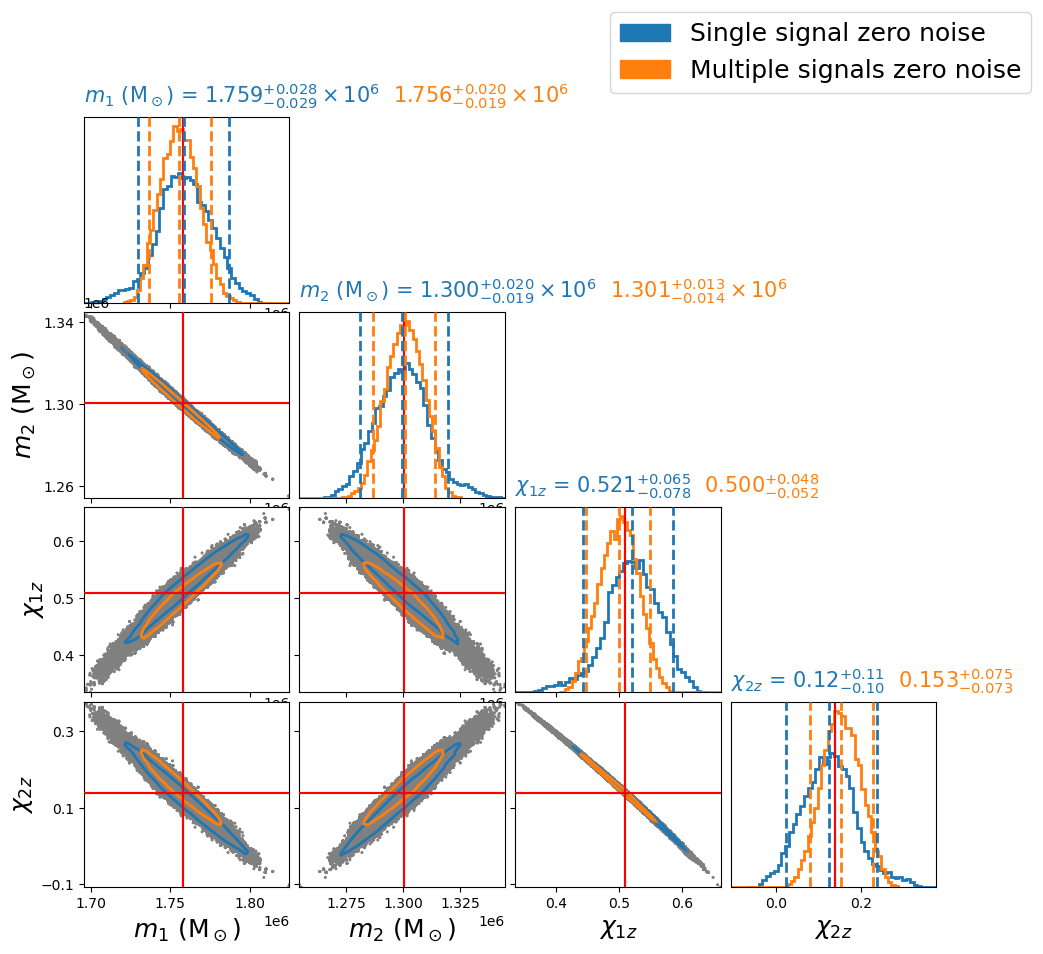"}.
	\caption{A corner plot showing the recovered intrinsic parameters posteriors of signal 0 in zero noise. The blue curves represent a year’s worth of data containing only signal 0. The orange curves represent data containing all 6 signals. The red lines represent the injected parameter values of the signal. The tiles at the top of each column display the median and $90\%$ range for each inference. The contour in the 2D tiles represents the $90\%$ confidence intervals.}
	\label{fig:int_zn}
\end{figure}

\noindent
When generating data with BBHx injections for the with and without GB cases, we used the Sangria training data set. The Gaussian LISA noise and GBs for Sangria blind are generated in the same way as the training data set, but are unique to the blind data set. The Sangria data has waveform injections that were created with the LDC code base and not injected with BBHx.

We will now present the results of each stage of this development. The parameters used to create the BBHx injections are identical to that of the signals in the Sangria blind data. We ran analysis on all signals injected with BBHx and Sangria blind, but will concentrate our discussion on the first signal (from time zero) which we refer to as signal 0. We will distinguish the Sangria blind injection of the first signal as \enquote{blind 0} as it is not produced with BBHx and is in the Sangria blind noise and GBs data. Quoted parameters found in the inference are the maximum log-likelihood sample from the posterior.

All of the BBHx inference runs were performed with the heterodyne parameters set to the true injected parameters. The final Sangria runs reference parameters were set to the parameters found in the search section of the pipeline. Figure \ref{fig:psd} shows the PSDs used for the first 3 stages before analysing the Sangria blind data. The same sampler settings are used for each analysis to ensure a fair comparison.

\begin{figure}[t!]
	\centering
	\includegraphics[width=1.0\linewidth]{"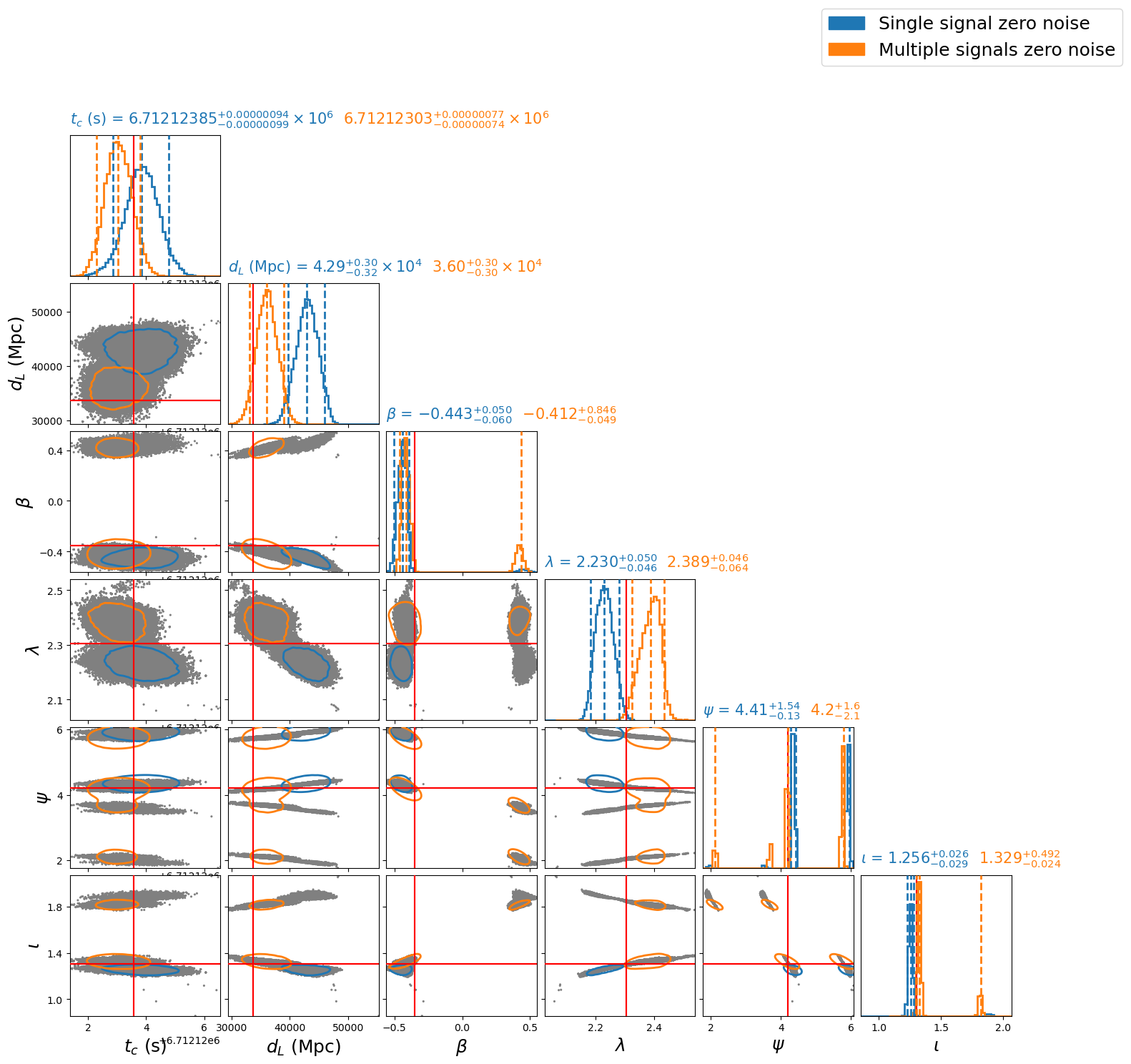"}.
	\caption{Extrinsic parameter posterior corner plot of injected BBHx signal in zero noise for signal 0. Red lines represent the injected parameter values of the signal. The tiles at the top of each column display the median and $90\%$ range for each inference. The contour in the 2D tiles represents the $90\%$ confidence intervals.}
	\label{fig:ext_zn}
\end{figure}

For each stage of development, we performed two sets of inference. One set of inference contained only signal 0 injected into the data, shown in blue in subsequent Figures, and the other contained all 6 signals from Sangria, shown in orange. All injections are created in a year's worth of data sampled at 5s intervals to match the convention of the Sangria challenge.

Note that all of these parameters are given in the detector frame. This is the standard return of BBHx. For the interest of those wanting to inspect parameters in the source frame, the redshift for signal 0 injections is $z=3.72$. Finally, we highlight the degeneracy, $h(\underline{\theta},\psi) = h(\underline{\theta}, \psi + \pi)$, which can be observed throughout the results section.

\subsubsection{Zero noise, BBHx injection -}

Figure \ref{fig:int_zn} presents the recovered intrinsic parameter posteriors for this analysis, and Figure \ref{fig:ext_zn} the extrinsic parameters, along with their $90\%$ confidence intervals. The black PSD curve from Figure \ref{fig:psd}, was used to scale the signals, to give SNRs of comparable sizes to subsequent analyses. We can see for the single and multiple injection cases that the intrinsic parameters are recovered within the $90\%$ confidence interval of the injected values. However, most extrinsic parameters do not fall within the $90\%$ confidence interval; the sky position $\beta$ has found the correct sky mode but the injected value falls just outside the peak of both single and multiple signal runs. The correct mode for $\lambda$ is also recovered, however, in both the single and multiple signal cases the $90\%$ confidence intervals do not capture the injected value.

\begin{figure}[t!]
	\centering
	\includegraphics[width=0.7\linewidth]{"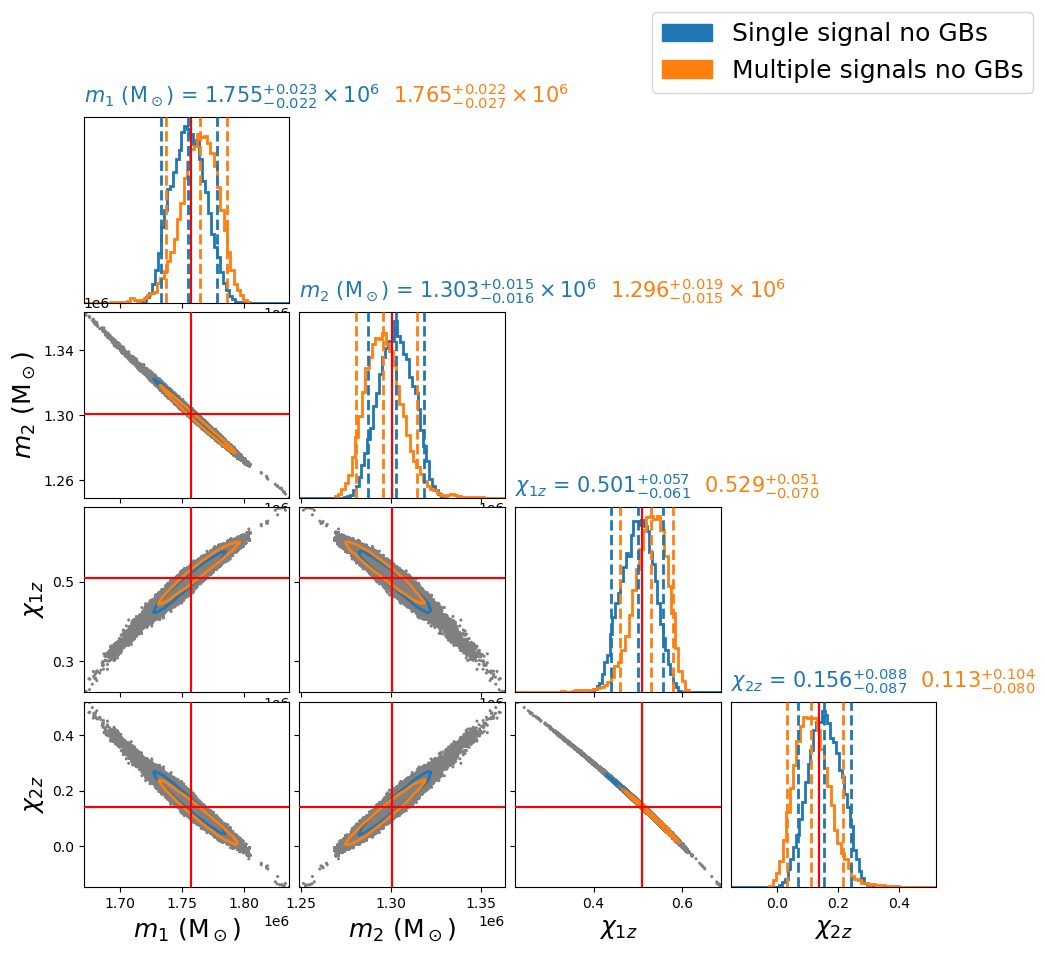"}.
	\caption{Intrinsic parameter posterior corner plot of injected BBHx signal(s) in Gaussian stationary noise without GBs for signal 0. Red lines represent the injected parameter values of the signal. The tiles at the top of each column display the median and $90\%$ range for each inference. The contour in the 2D tiles represents the $90\%$ confidence intervals.}
	\label{fig:int_allsig_withoutgbs}
\end{figure}

\subsubsection{Sangria noise without GBs, BBHx injections -}These signals are injected into Gaussian, stationary noise using the Sangria training data. The black curve in Figure \ref{fig:psd} is the PSD used for this stage.

\begin{figure}[t!]
	\centering
	\includegraphics[width=1.0\linewidth]{"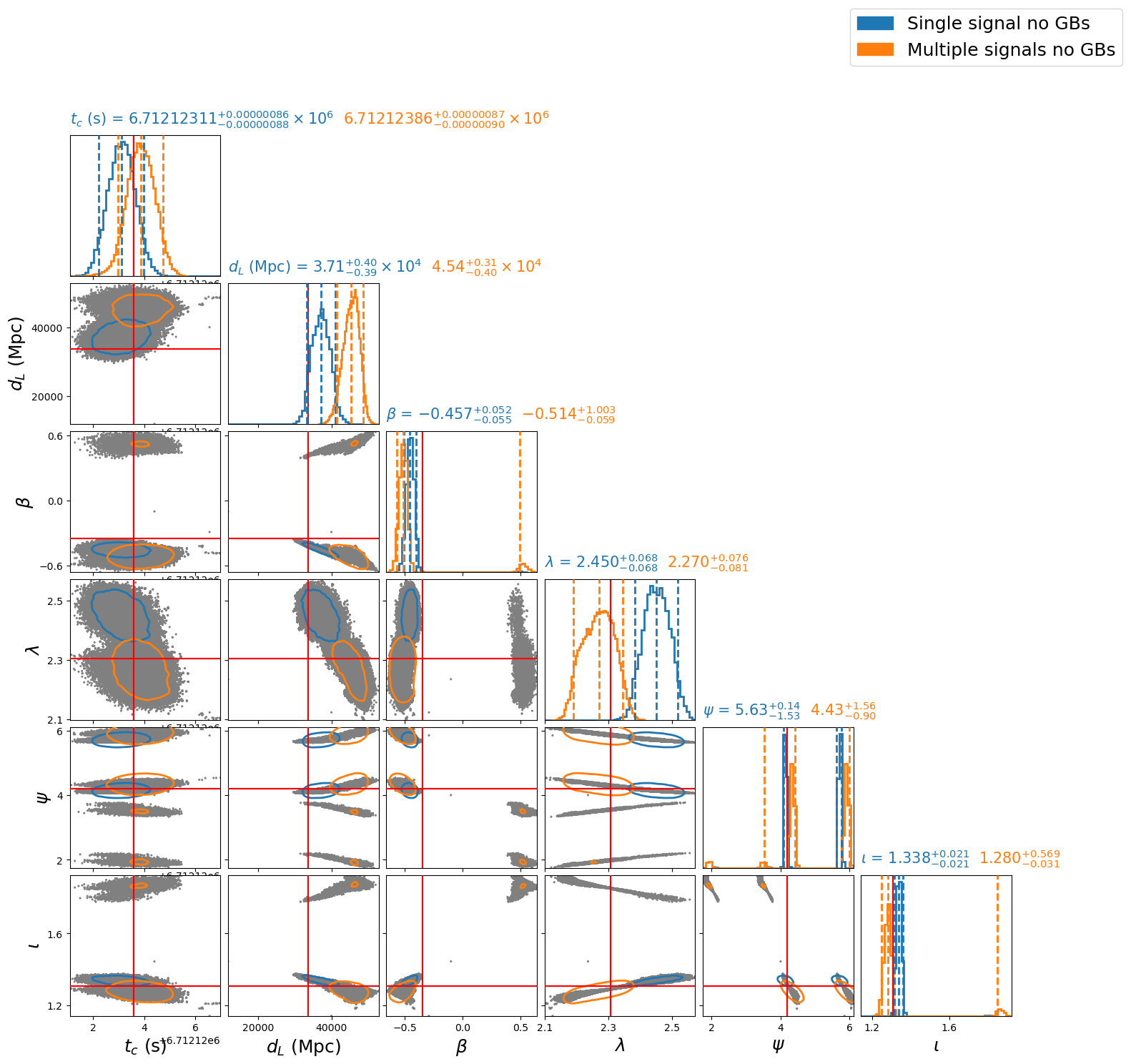"}.
	\caption{Extrinsic parameter posterior corner plot of injected BBHx signal(s) in Gaussian stationary noise without GBs for signal 0. Red lines represent the injected parameter values of the signal. The tiles at the top of each column display the median and $90\%$ range for each inference. The contour in the 2D tiles represents the $90\%$ confidence intervals.}
	\label{fig:ext_allsig_withoutgbs}
\end{figure}

As Figure \ref{fig:int_allsig_withoutgbs} shows, the inclusion of Gaussian noise when comparing to the zero noise case for both multiple and single injections has no negative bias to the inferred intrinsic parameters. All intrinsic parameters fall within the $90\%$ confidence interval in both cases. The extrinsic parameter recovery, shown in Figure \ref{fig:ext_allsig_withoutgbs}, shows no notable differences when comparing to the zero noise case. We do start to see some preference to other likelihood modes in sky position as in two cases (signals 2 and 4) the inference preferred the other high likelihood mode of $\beta$ instead of the true injected mode. This is true in both the single injection case and the multiple injection case.

\subsubsection{Sangria noise with GBs, BBHx injections -}The red curve in Figure \ref{fig:psd} is the PSD used for this stage and contains all of the noise from the GBs. We see that this significantly changes the shape of the PSD in the $10^{-3}-10^{-2}Hz$ region of frequency space.

\begin{figure}[t!]
	\centering
	\includegraphics[width=0.7\linewidth]{"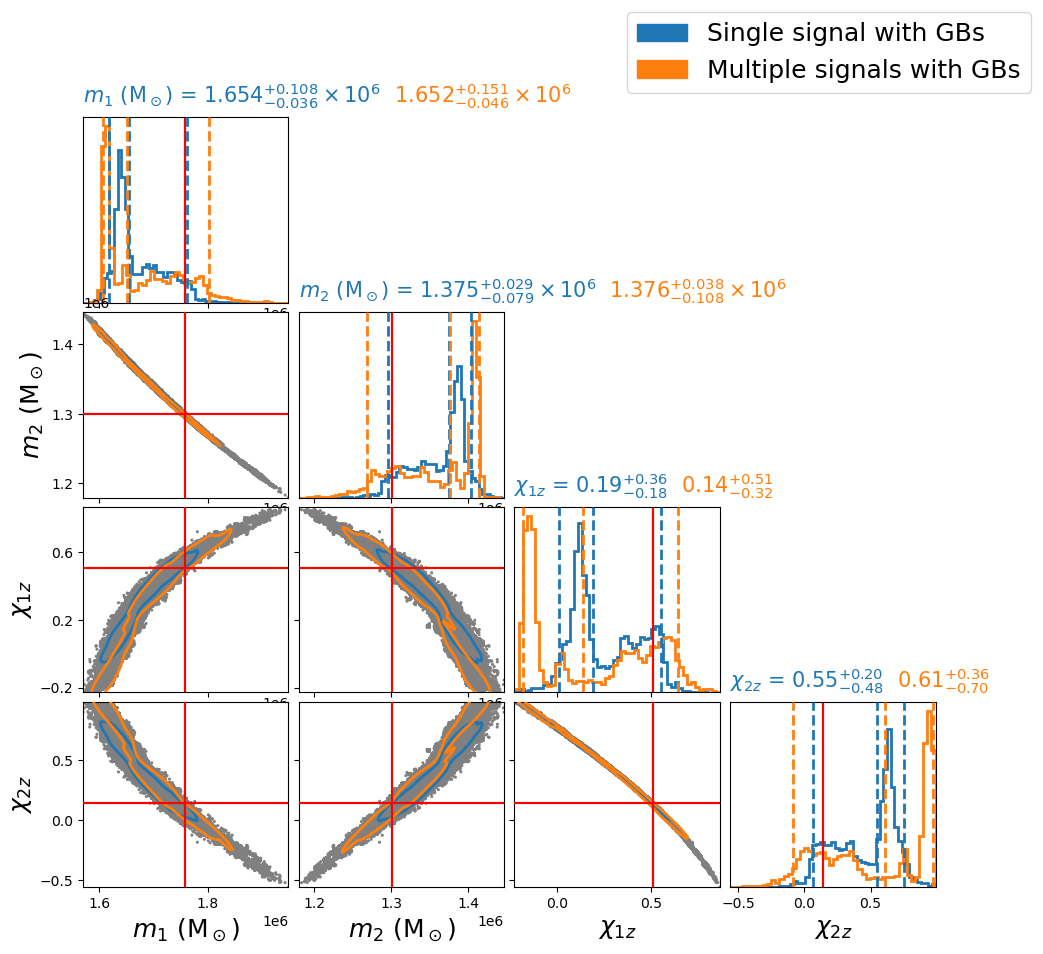"}.
	\caption{Intrinsic parameter posterior corner plot of injected BBHx signal(s) in Gaussian stationary noise with GBs for signal 0. Red lines represent the injected parameter values of the signal. The tiles at the top of each column display the median and $90\%$ range for each inference. The contour in the 2D tiles represents the $90\%$ confidence intervals.}
	\label{fig:int_allsig_withgbs}
\end{figure}

\begin{figure}[t!]
	\centering
	\includegraphics[width=1.0\linewidth]{"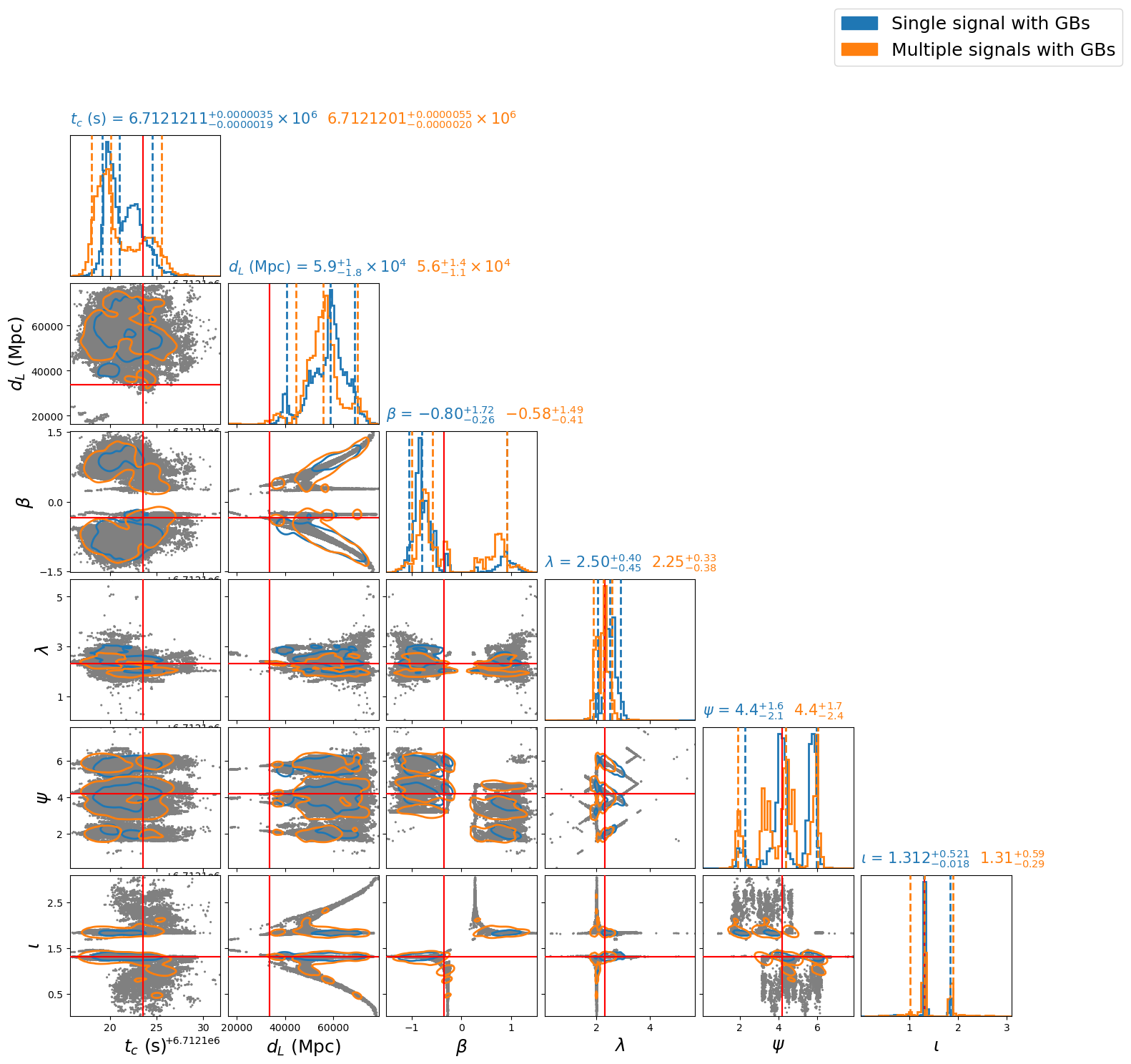"}.
	\caption{Extrinsic parameter posterior corner plot of injected BBHx signal(s) in Gaussian stationary noise with GBs for signal 0. Red lines represent the injected parameter values of the signal. The tiles at the top of each column display the median and $90\%$ range for each inference. The contour in the 2D tiles represents the $90\%$ confidence intervals.}
	\label{fig:ext_allsig_withgbs}
\end{figure}

The posteriors in Figures \ref{fig:int_allsig_withgbs} and \ref{fig:ext_allsig_withgbs} are broader which is expected given the noise curve with GBs present. Figure \ref{fig:int_allsig_withgbs} demonstrates that the intrinsic parameters still fall within the $90\%$ confidence interval.

\begin{table}[b!]
    \label{p_val_test}
    \begin{center}
        \begin{tabular}{||c|c|c|c||}
        \hline
            Signal & Sangria noise without GBs & Sangria noise with GBs & Sangria blind \\\hline 
            0 & 78\% & 34\% & 93\% \\\hline
            1 & 41\% & 84\% & 98\% \\\hline
            2 & 20\% & 44\% & 99\% \\\hline
            3 & 2.7\% & 84\% & 100\% \\\hline
            4 & 36\% & 84\% & 81\% \\\hline
            5 & 18\% & 6.7\% & 86\% \\\hline
        \end{tabular}
    \end{center}
    \caption{Each percentage represents the amount of samples whose likelihood value is greater than the likelihood of the corresponding injected waveform. } 
\end{table}

GBs have had the most significant effect on inference thus far as the peaks in the posterior likelihood have shifted from the injected value when compared to the no GBs case. We also see a lot more structure in the posteriors that includes GBs. In Table 5 we have calculated the percentage of samples from the posterior with a higher likelihood than that of the likelihood of the injected waveform. If the assumptions of Gaussianity and stationarity are valid, these values for a collection of posteriors from different signals/injections should be uniformly distributed. We see for the case of Gaussian stationary noise without GBs that the distribution of values does appear to be uniform. The test performed here would have benefited from $\mathcal{O}(100)$ posteriors to properly conclude that the distribution is uniform. When including GBs, we can see that the distribution is no longer uniform. The effect of GBs being non-stationary therefore has biased the inferred parameters away from the injected value. Even with only six signals we can clearly see the bias when the data contains GBs and when inspecting the full Sangria data. We have presented the other 5 signal parameter injections in Appendix A. We can clearly see that Figures A1-A5 show a bias in intrinsic parameter recovery as signal 0.

As for the extrinsic parameters for signal 0, the inferred values represent the injected waveform well. This is not true for the other signals in the data as most struggle to find the correct sky mode for $\beta$ and/or $\lambda$. 

\begin{table}[t!]
\label{diff_inf_0}
\hspace{2cm}
\begin{tabular}{||l|c|c||}
    \hline
    \multicolumn{3}{||c||}{\textbf{Blind 0}} \\
    \hline
    \bfseries Parameters & \bfseries Injected value & \bfseries Recovered value 
    \begin{filecontents*}{"diff_table_inf_01.csv"}
Parameters,Injected value,Recovered value
m_1~(M_{\odot}),1.76 \times 10^{6},1.70^{+0.13}_{-0.11} \times 10^{6}
m_2~(M_{\odot}),1.30 \times 10^{6},1.34^{+0.083}_{-0.089} \times 10^{6}
a_1,0.51,0.39^{+0.32}_{-0.56}
a_2,0.14,0.29^{+0.66}_{-0.49}
t_c~(s),6712124,6712466^{+103}_{-793}
\iota~(rad),1.31,1.32^{+0.52}_{-0.02}
\psi~(rad),1.06,1.44^{+1.03}_{-1.08}
\beta~(rad),-0.35,-0.05^{+0.77}_{-0.59}
\lambda~(rad),2.31,2.11^{+2.40}_{-0.79}
\end{filecontents*}
\csvreader{"diff_table_inf_01.csv"}{}
    {\\\hline$\csvcoli$&$\csvcolii$&$\csvcoliii$}
    \\\hline
\end{tabular}
\caption{The injected and recovered parameters for signal 0 in the blind Sangria dataset. The recovered values are stated with the 90\% confidence regions. }
\end{table}

\begin{figure}[b!]
	\centering
	\includegraphics[width=0.7\linewidth]{"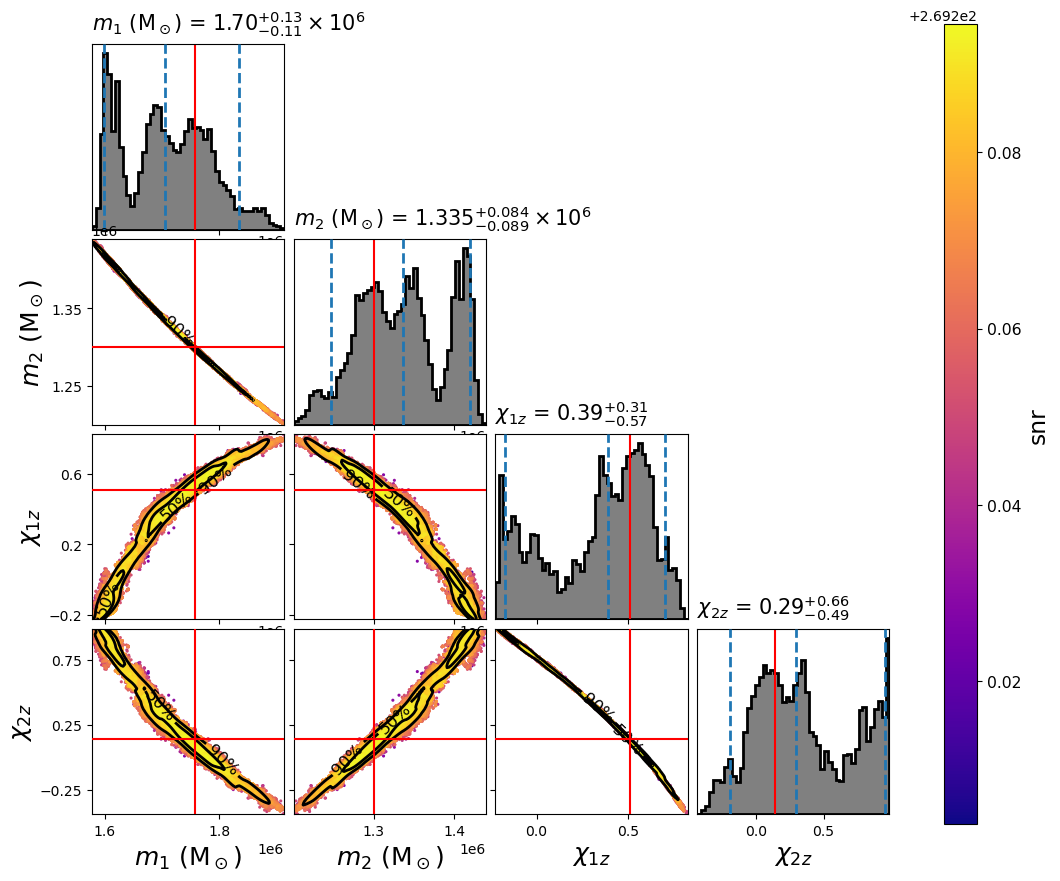"}.
	\caption{Intrinsic parameter posterior corner plot of Sangria blind data for signal 0. Red lines represent the injected parameter values of the signal. The tiles at the top of each column display the median and $90\%$ range for each inference. The contour in the 2D tiles represents the $90\%$ confidence intervals.}
	\label{fig:int_sangria}
\end{figure}

\subsubsection{Sangria blind -}The full parameter recovery of blind 0 in Sangria is presented in Table 6, which also displays the injected parameter values. The intrinsic parameters in Figure \ref{fig:int_sangria} fall within the $90\%$ confidence interval. However, this is not true for all signals in the data set. There is one case where all intrinsic parameters peak at roughly the $88\%$ confidence interval. Overall, the peaks in likelihood are far less prominent when compared to the previous cases.

Extrinsic parameters also have far less prominent peaks than previous cases and suffer from the same sky position degeneracy issues where the wrong mode is commonly favoured. For half of the cases, $D_L$ is recovered within the $90\%$ interval and for all cases, $t_c$ and $\iota$ are recovered in the $90\%$ confidence interval.

\begin{figure}[t!]
	\centering
	\includegraphics[width=1.0\linewidth]{"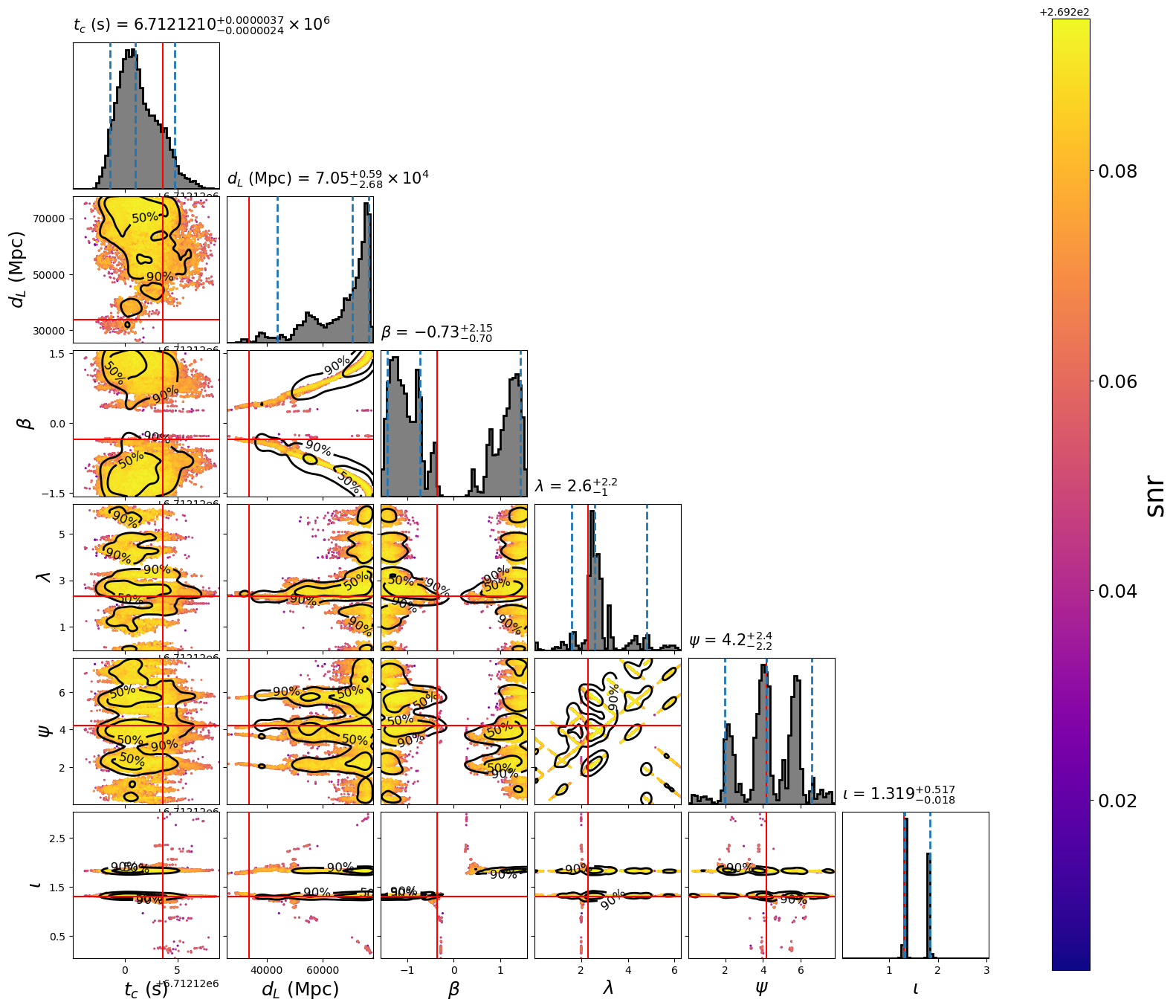"}.
	\caption{Extrinsic parameter posterior corner plot of Sangria blind data for signal 0. Red lines represent the injected parameter values of the signal. The tiles at the top of each column display the median and $90\%$ range for each inference. The contour in the 2D tiles represents the $90\%$ confidence intervals.}
	\label{fig:ext_sangria}
\end{figure}

Figure \ref{fig:ext_sangria} shows the extrinsic recovery of our reference signal and is the only example out of the six signals that recovers the correct mode for $\lambda$. Extrinsic parameter recovery would be greatly improved when analysing signals with higher order modes \cite{Marsat_2021}. As this data challenge uses just (2,2) mode injections, it was expected that the extrinsic parameters would be the most difficult to recover correctly. 

\newpage

\section{Conclusion}

The application of a template bank search was able to identify all MBHB signals in the Sangria mock data challenge. Although the search for MBHB signals is relatively trivial due to the high SNR, being able to provide reference parameters for the heterodyne likelihood is the main advantage when taking this approach. We can produce a template bank $\mathcal{O}(minutes)$ to fulfill a minimal match of 0.9. Given that the SNR recovered is greater than $92\%$, we find that this technique is a viable approach to quickly search through LISA data for MBHBs. At the same time, the templates returned from the search provide appropriately high likelihood reference parameters for heterodyning which do not bias parameter estimation. To expand and improve upon the search, we could consider a coincident search over both the A and E TDI data streams. This would filter each TDI stream separately and then check for coincident triggers.

We have shown that the introduction of different forms of noise can have an effect on both intrinsic and extrinsic parameter recovery, although the extrinsic parameter recovery suffers more than the intrinsic. Something to consider about this data set is how the injected signals only contain the dominant (2,2) mode. Other works have shown that sky position recovery is vastly improved as other sub dominant modes are introduced to parameter estimation \cite{Marsat_2021}. A further study could be to investigate the inference of a multi-modal signal, in the presence of LISA instrumental noise. We could then go further by introducing non-stationarities in the data, with the inclusion of GBs. 

The intrinsic parameter recovery of MBHBs in the presence of confusion noise caused by GBs can bias the result away from the true value. Although the injected values in our work did fall within the $90\%$ confidence intervals, the peak likelihood has moved away from the injected value when compared to the no GB inference case, see Figures \ref{fig:int_allsig_withoutgbs} and \ref{fig:int_allsig_withgbs}. This motivates the need for a full global fit pipeline, that can identify and infer the properties of the GBs and then remove them from the data before running inference on MBHBs. This was first described in Ref \cite{Cornish:2005qw}. 

An area of further investigation concerns how we estimate PSDs. As stated previously, in this work we estimate our PSDs by reading in the entire year's worth of data. Due to LISA's motion across the sky, this makes the PSDs non-stationary. One possible method is to construct our PSD by assuming quasi-stationarity on smaller time chunks of the year's worth of data, and assemble these into one PSD using the time/frequency track of the reference signal in question. It would also be interesting to investigate what effects the PSD being non-stationary has on the search and inference of such high SNR signals.

\newpage

\section{Acknowledgements}
We would like to thank Stas Babak, Quentin Baghi, Maude Le Jeune, Jean-Baptiste Bayle, and Natalia Korsakova for helpful discussions regarding LISA data analysis. We also thank Michael Katz for help with using BBHx and for making his code publicly available for use. We also thank the GW group at the University of Portsmouth for their discussion and support during the project. CW would like to particularly thank Gareth Cabourn Davies, Arthur Tolley and Ellie Wadman for their advice, feedback and support. We are grateful to the computing teams from the University of Portsmouth and AEI Hannover for their significant technical support. CW is supported by a STFC studentship. LKN acknowledges support from the UKRI Future Leaders Fellowship through grant MR/T01881X/1. IWH and LKN acknowledge support from the UK Space Agency through grant ST/X002225/1. Numerical computations were done on the Sciama High Performance Compute (HPC) cluster which is supported by the ICG, SEPNet and the University of Portsmouth.

\section*{References}

\addcontentsline{toc}{section}{References}
\bibliographystyle{iopart-num}
\bibliography{main.bib}

\appendix

\section{Intrinsic parameter inference plots with GBs}

\begin{figure}[h!]
	\centering
	\includegraphics[width=0.7\linewidth]{"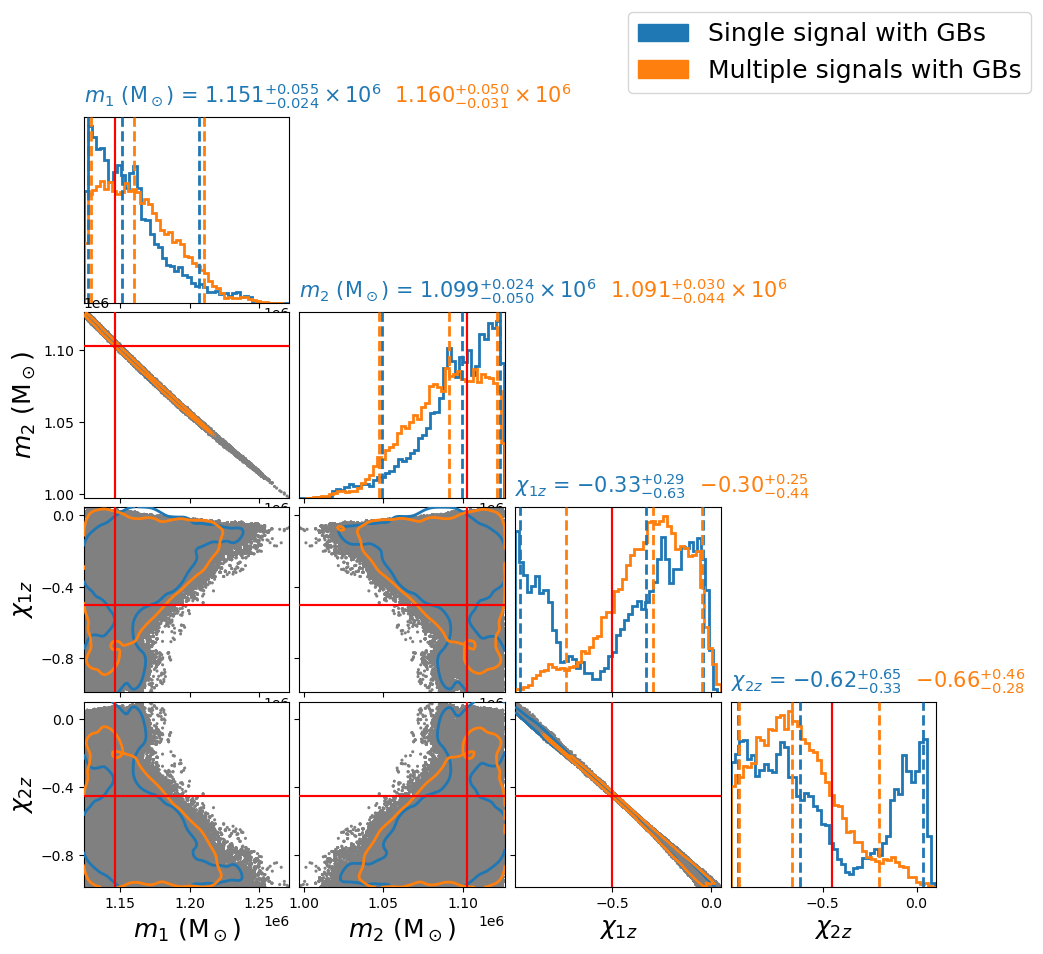"}.
	\caption{Inference for BBHx injection of signal 1 from Sangria blind over intrinsic parameters.}
	\label{fig:a1}
\end{figure}

\begin{figure}[h!]
	\centering
	\includegraphics[width=0.7\linewidth]{"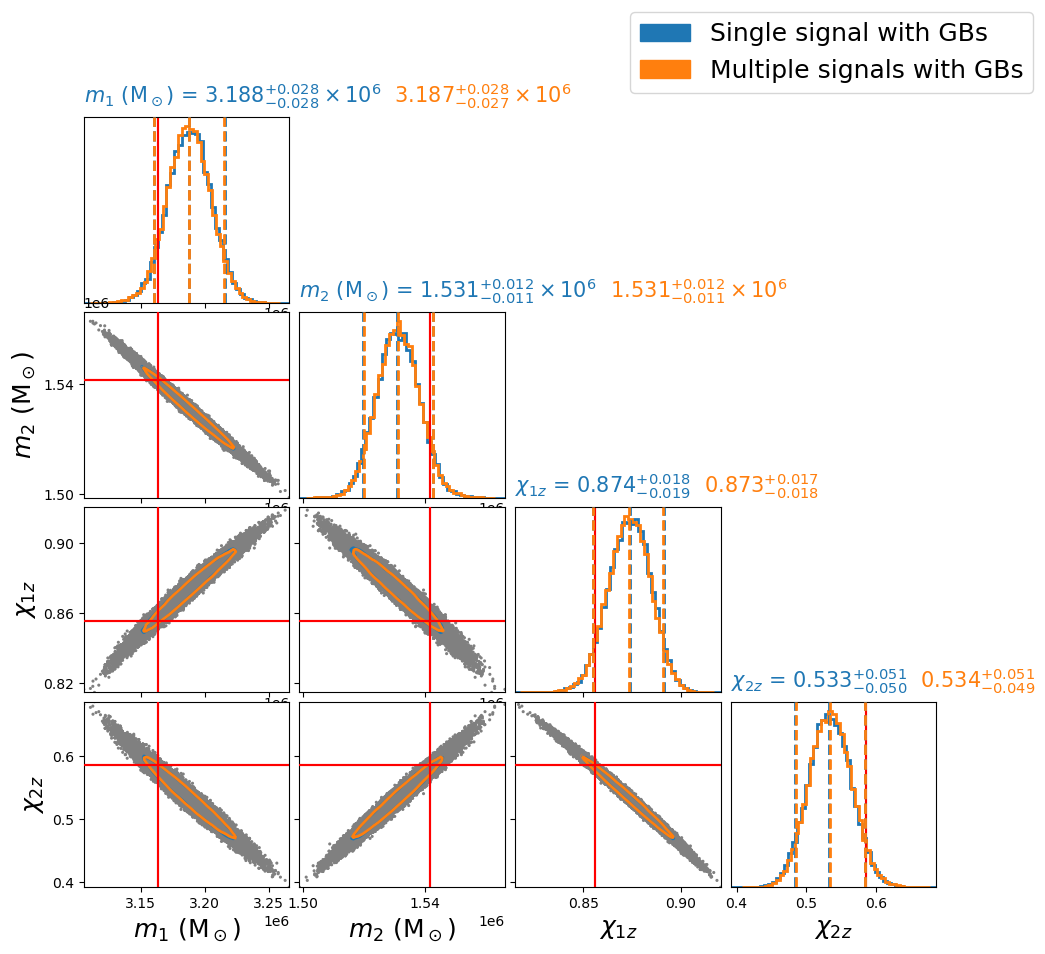"}.
	\caption{Inference for BBHx injection of signal 2 from Sangria blind over intrinsic parameters.}
	\label{fig:a2}
\end{figure}

\begin{figure}[h!]
	\centering
	\includegraphics[width=0.7\linewidth]{"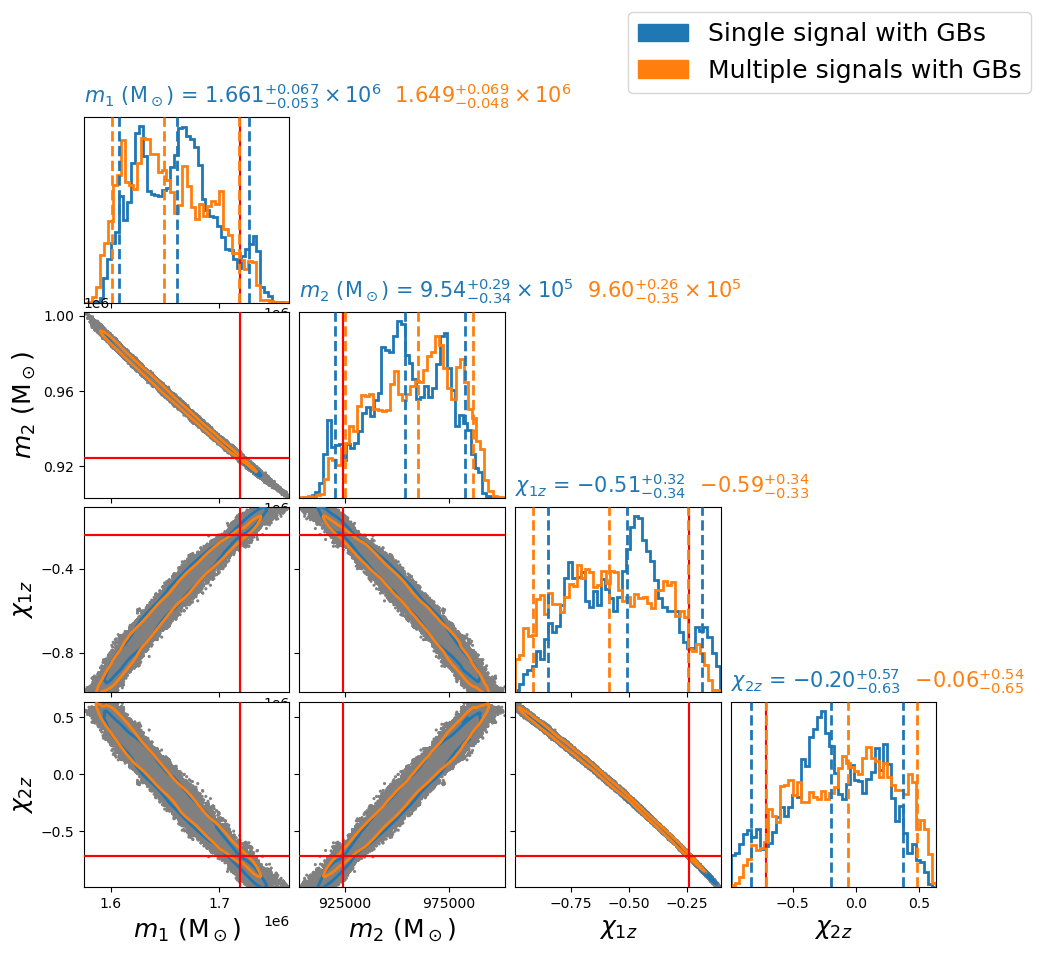"}.
	\caption{Inference for BBHx injection of signal 3 from Sangria blind over intrinsic parameters.}
	\label{fig:a3}
\end{figure}

\begin{figure}[h!]
	\centering
	\includegraphics[width=0.7\linewidth]{"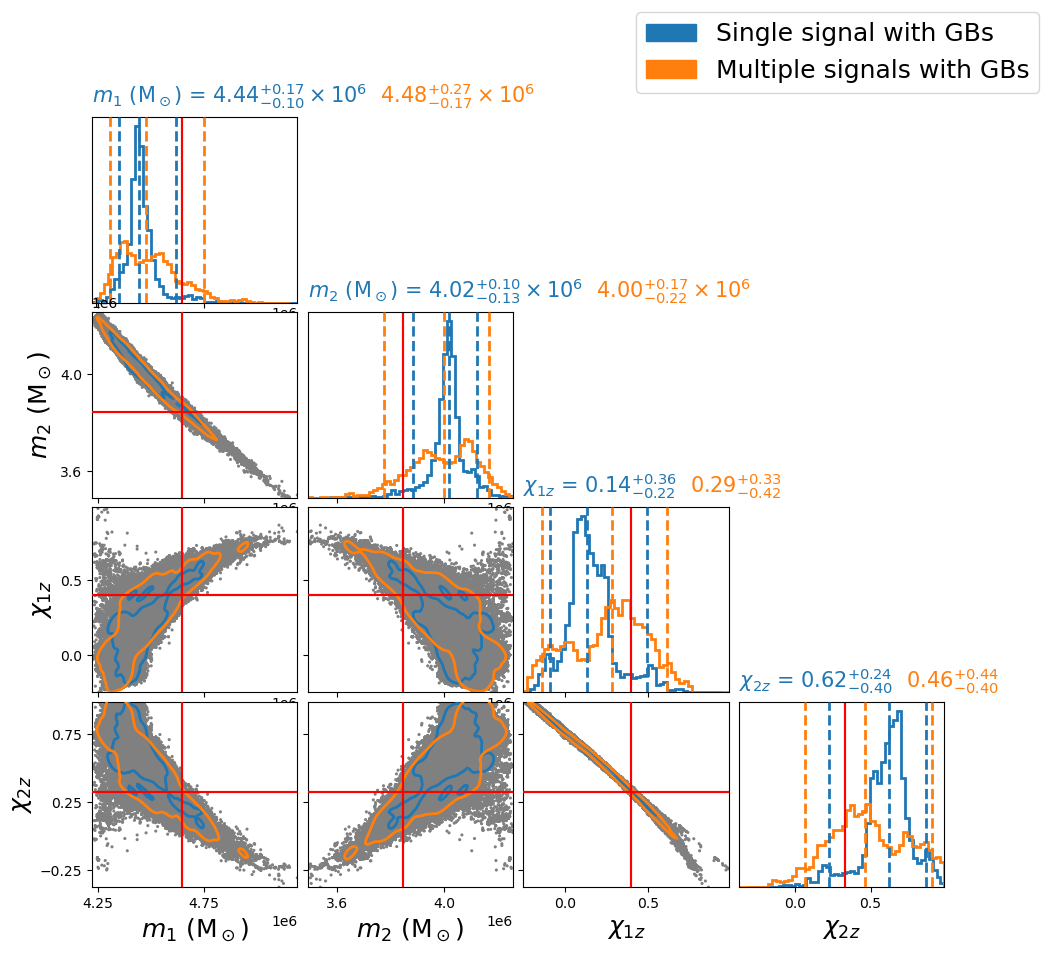"}.
	\caption{Inference for BBHx injection of signal 4 from Sangria blind over intrinsic parameters.}
	\label{fig:a4}
\end{figure}

\begin{figure}[h!]
	\centering
	\includegraphics[width=0.7\linewidth]{"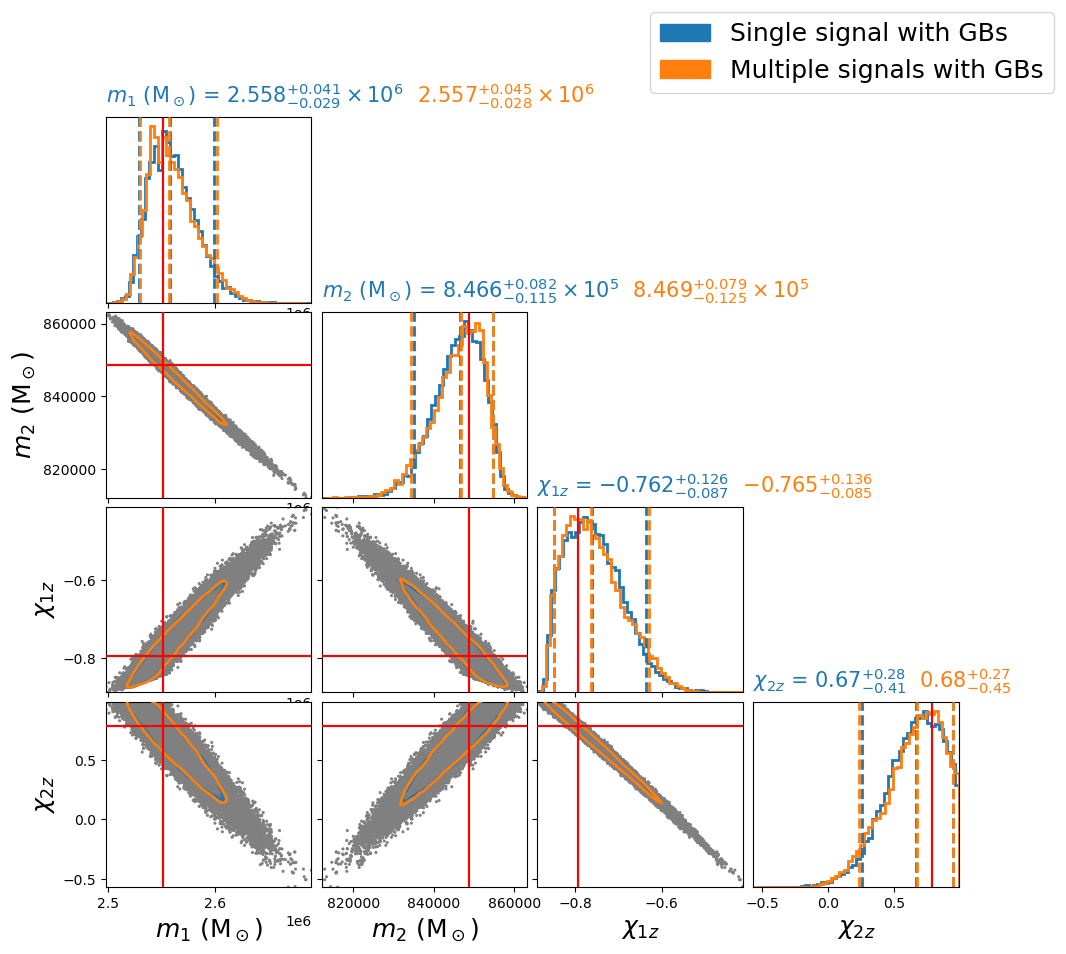"}.
	\caption{Inference for BBHx injection of signal 5 from Sangria blind over intrinsic parameters.}
	\label{fig:a5}
\end{figure}

\end{document}